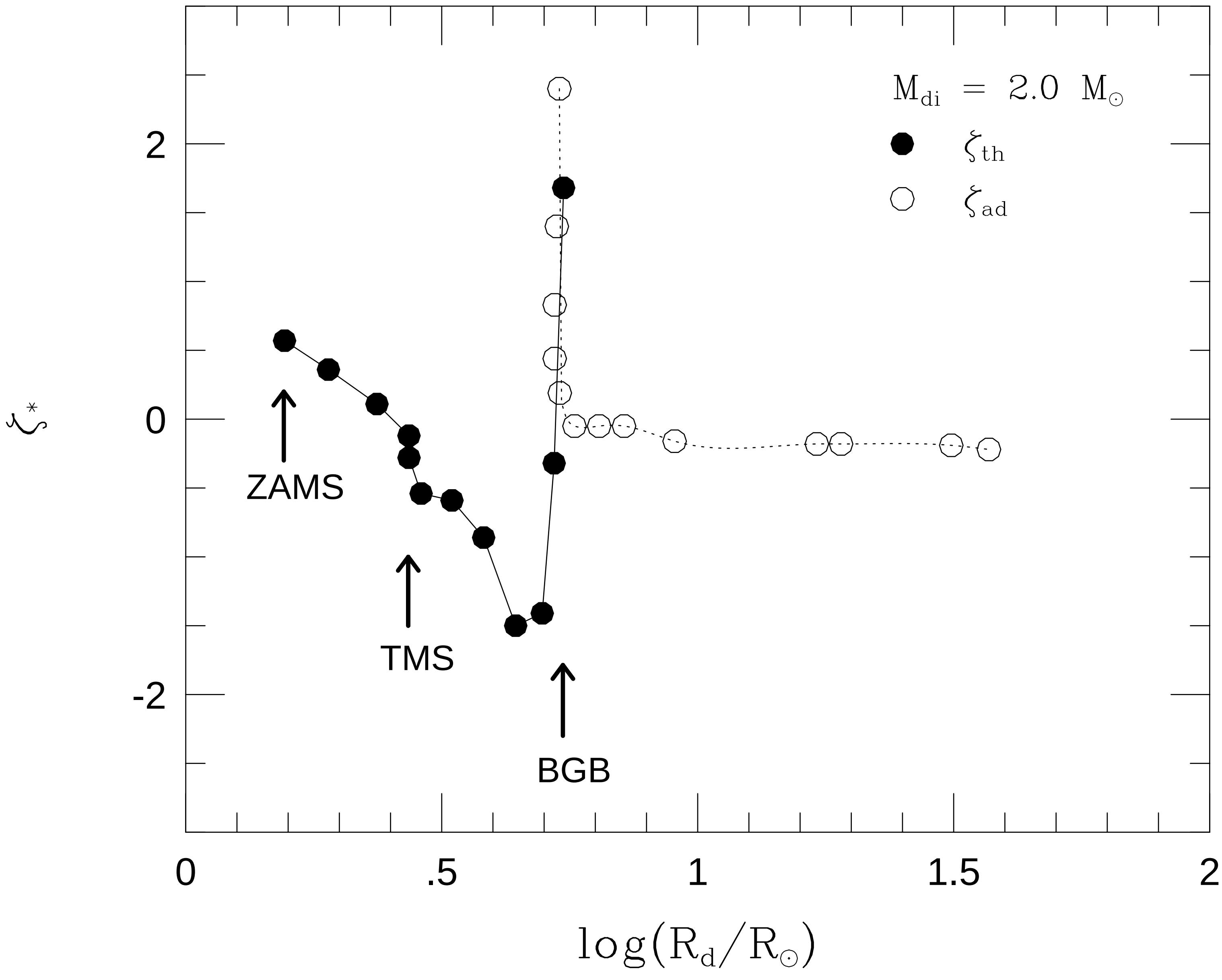

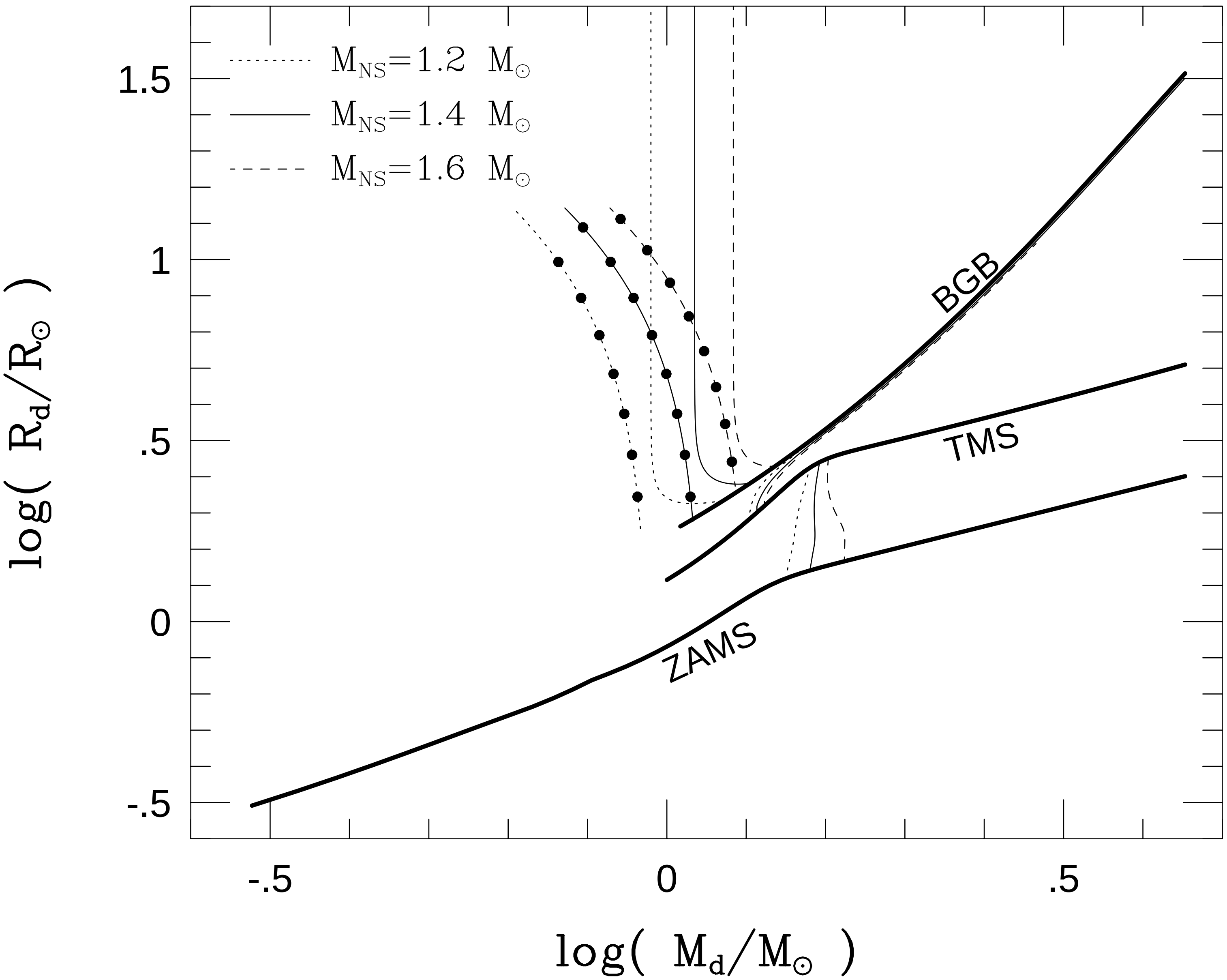

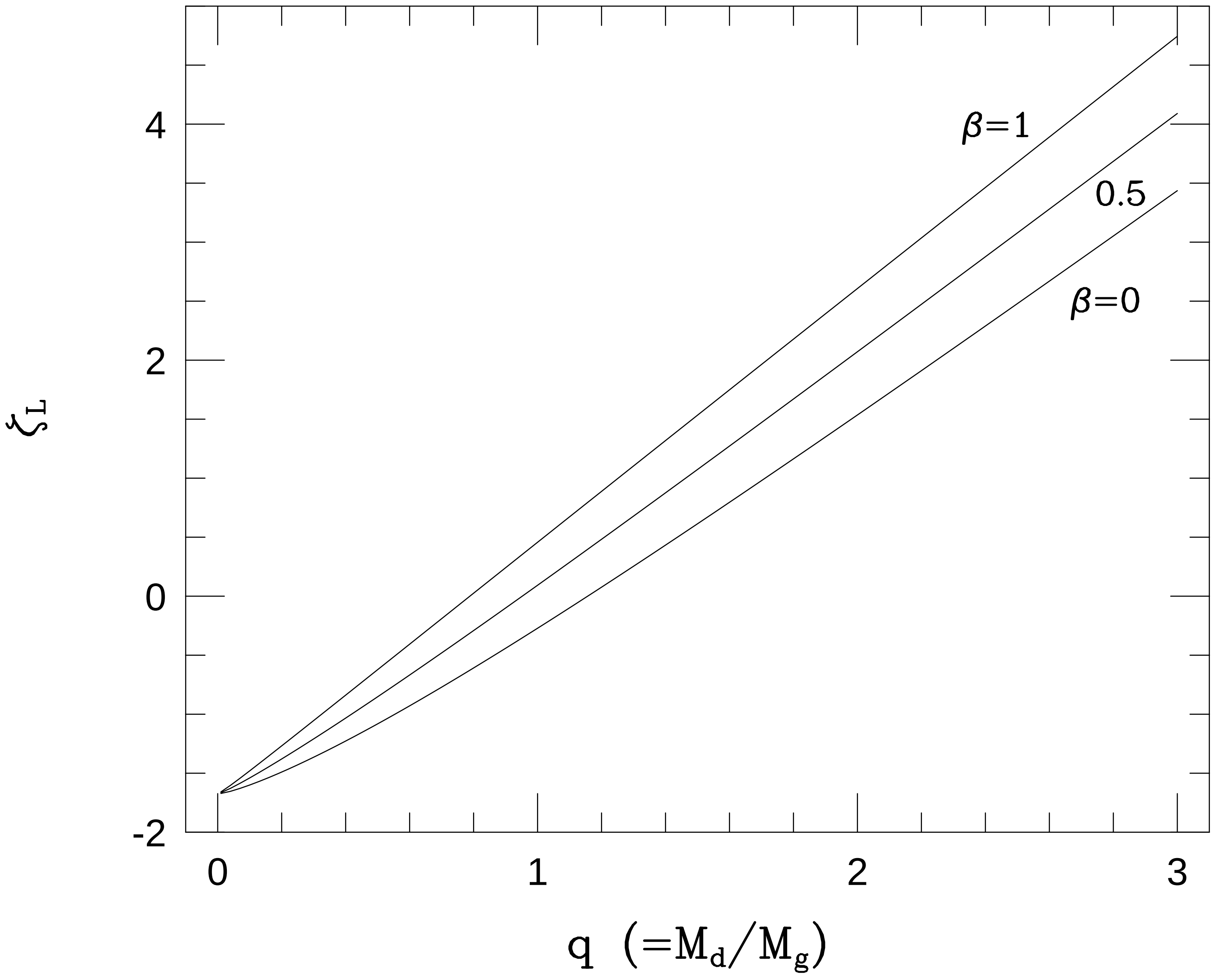

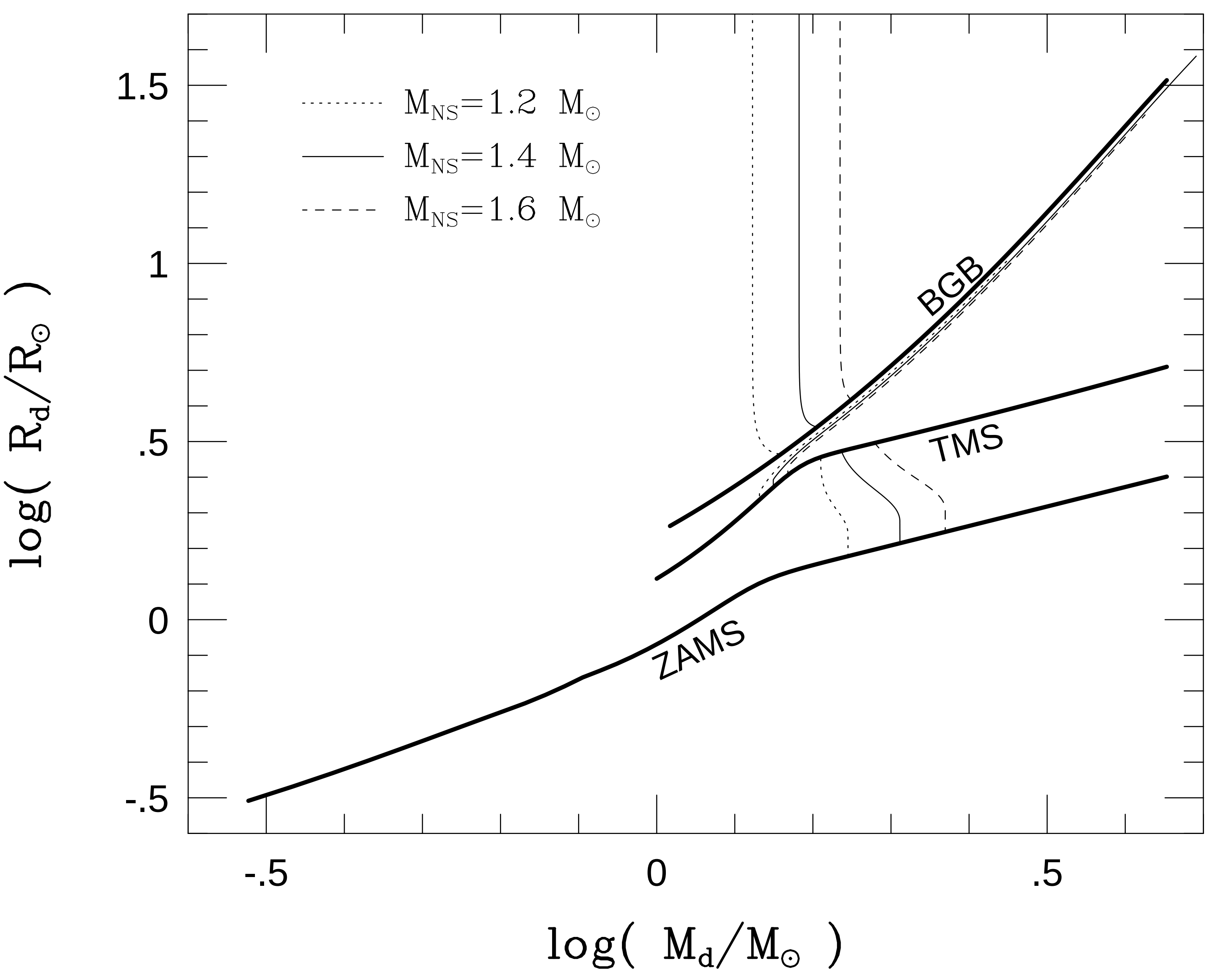

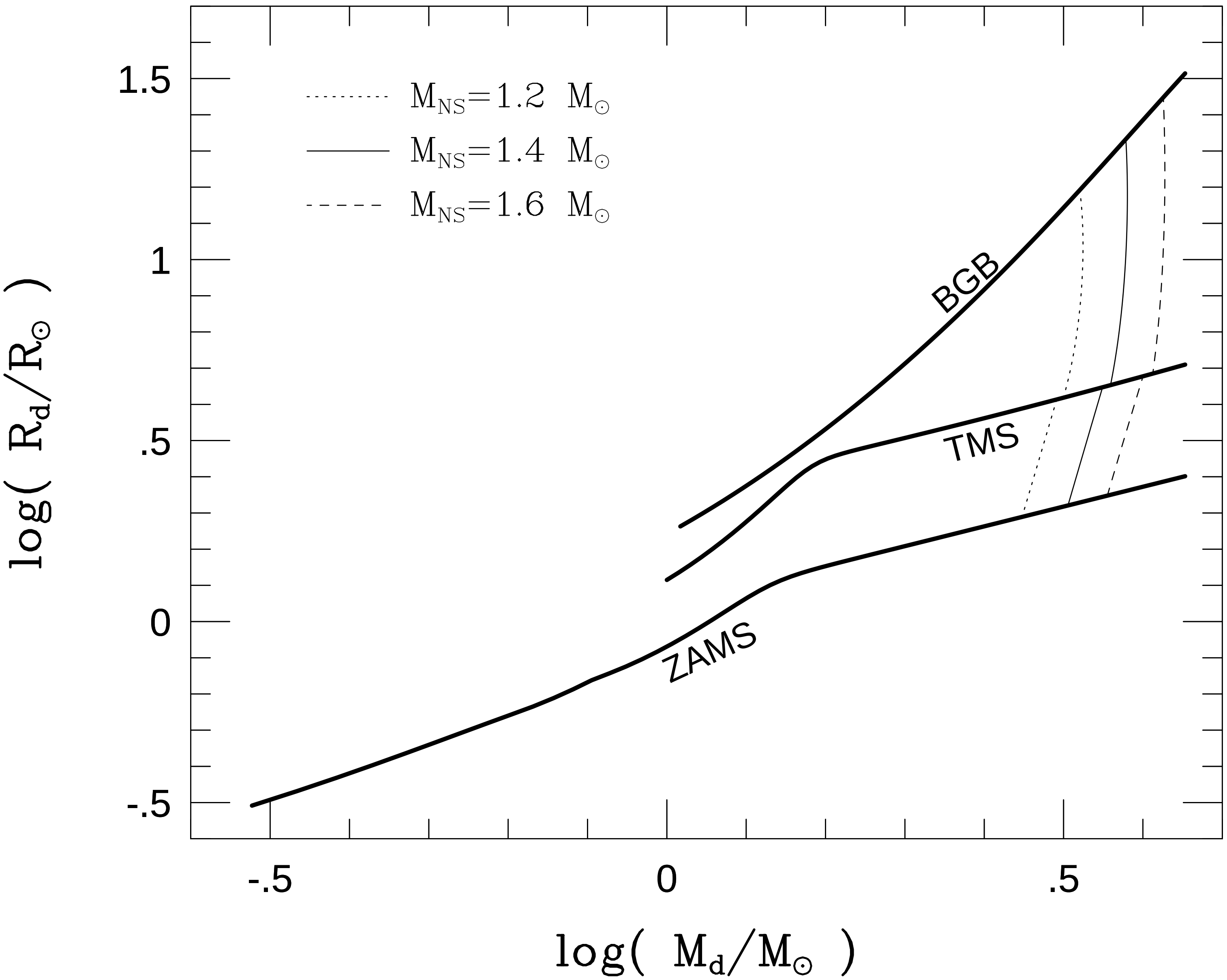

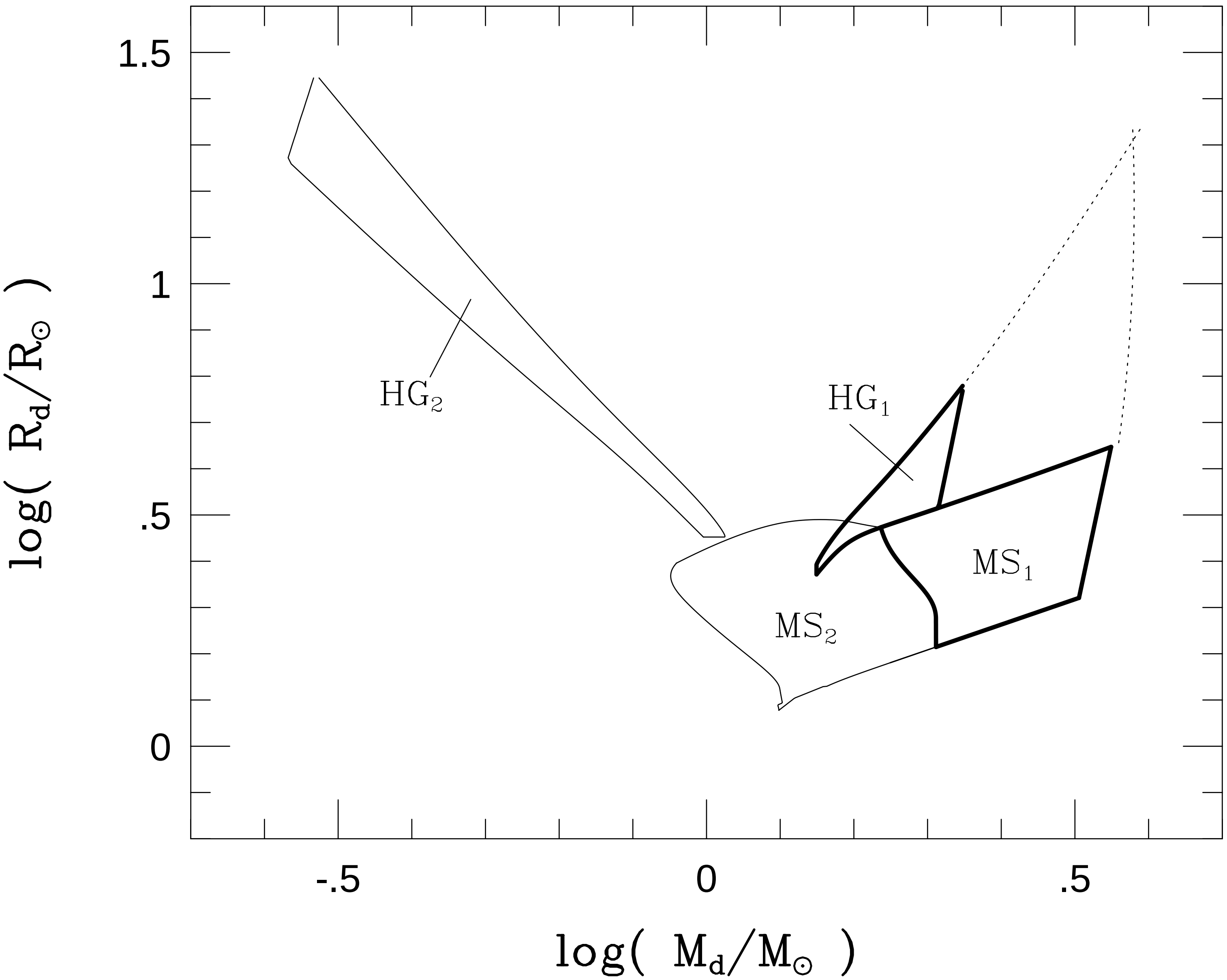

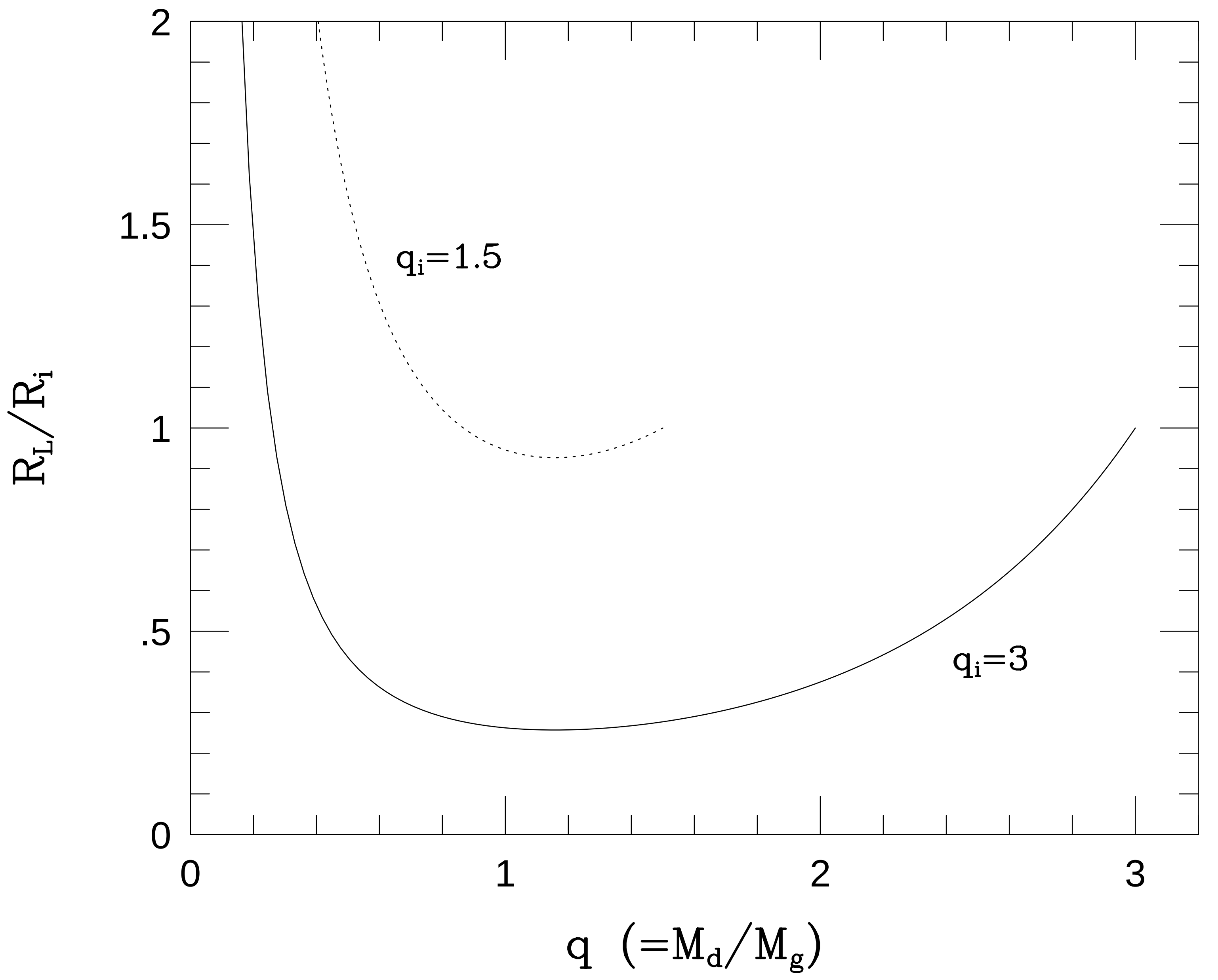

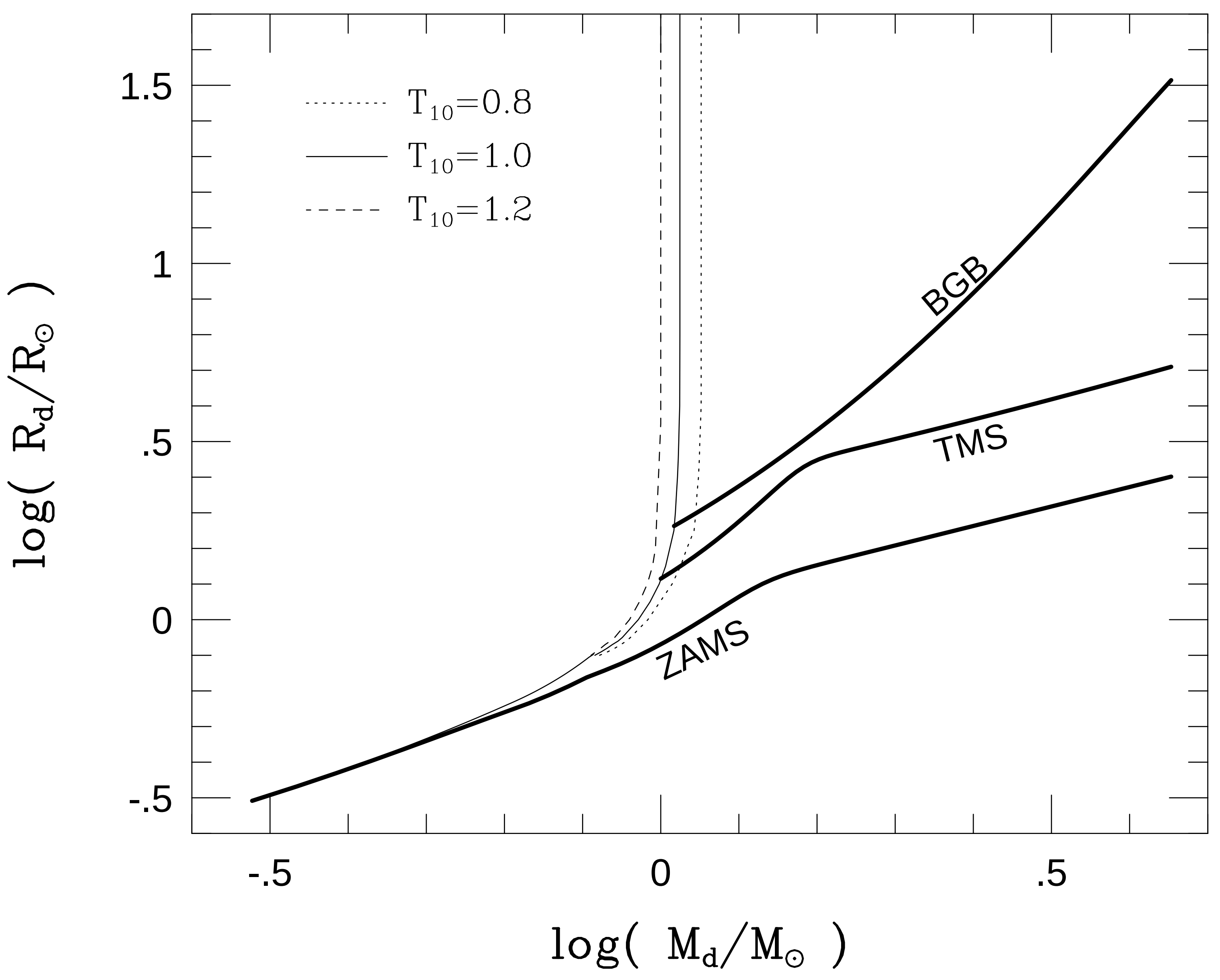

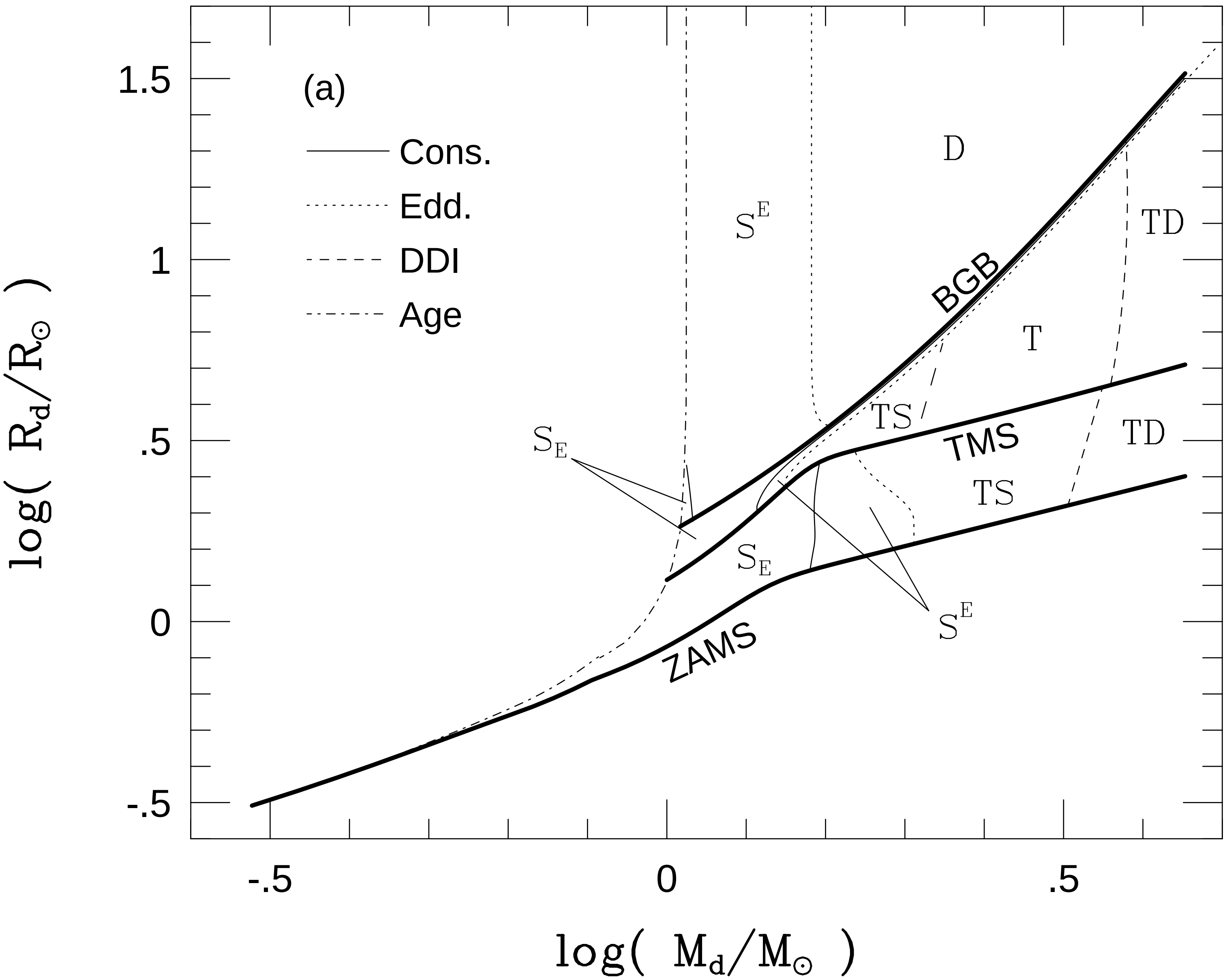

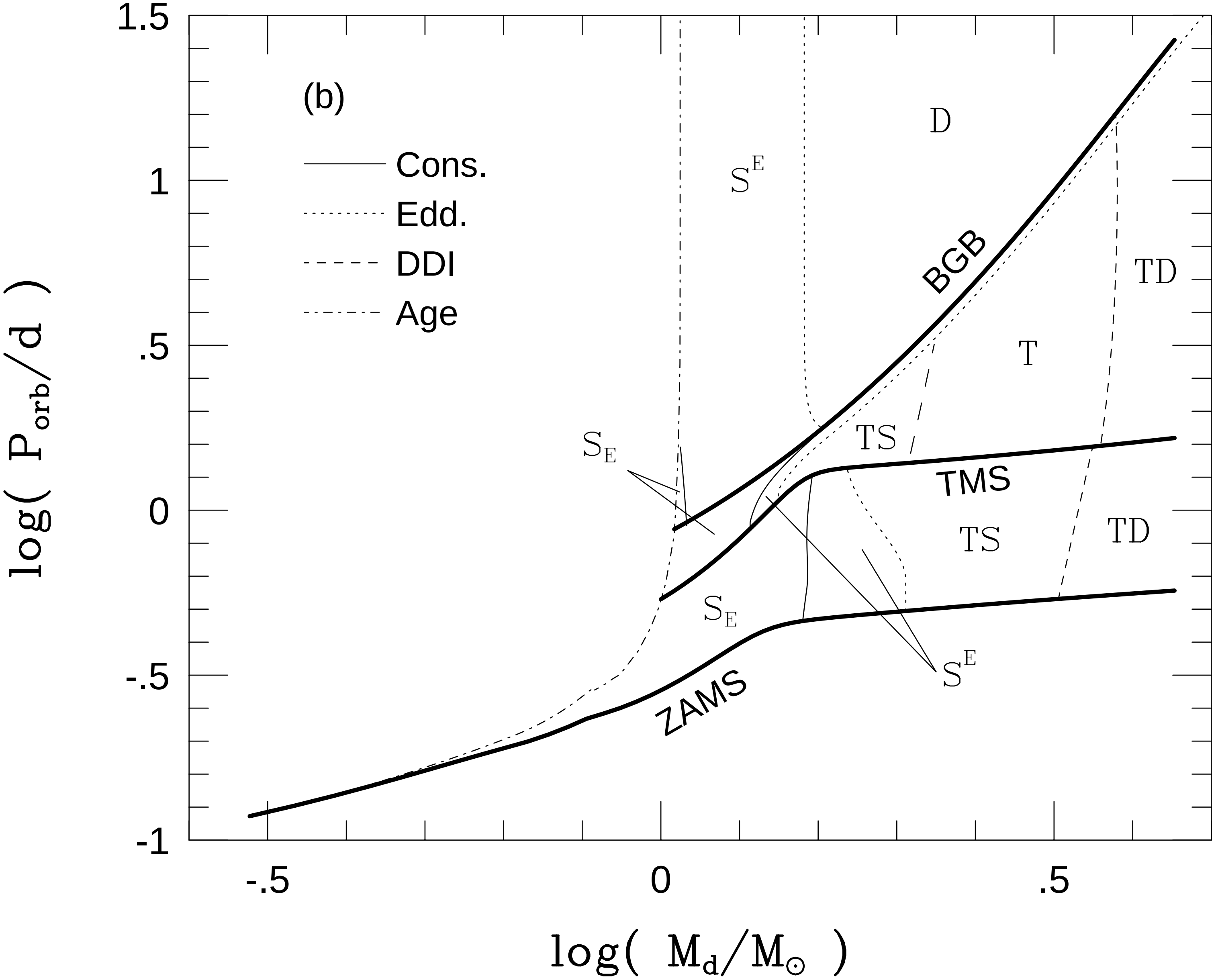

# Formation of Low-Mass X-ray Binaries. I. Constraints on Hydrogen-Rich Donors at the Onset of the X-ray Phase

Vassiliki Kalogera and Ronald F. Webbink
Astronomy Department, University of Illinois at Urbana-Champaign,
1002 West Green St., Urbana, IL 61801.
e-mail: vicky, webbink@astro.uiuc.edu

## ABSTRACT

We identify and quantify the set of constraints that neutron star-normal star binaries must satisfy in order to become observable LMXBs. These constraints are related to (i) the thermal and hydrostatic equilibrium of the donors, (ii) the degree to which the mass transfer process is conservative, and (iii) the age of the systems. They divide the parameter space of potential LMXBs in several distinct parts, of which those that actually become LMXBs at the onset of mass transfer occupy only a small part. Of the remainder, many become unstable to dynamical time scale mass transfer either at the onset or later in the course of mass transfer, and enter common envelope evolution. Others experience super-Eddington mass transfer but may eventually survive to become LMXBs. These survivors arguably include binary millisecond pulsars with orbital periods in excess of 100 d, ultrashort-period LMXBs with hydrogen-deficient donors, and long-period LMXBs with giant donors.

## 1. INTRODUCTION

A distinct feature of binary star systems is the possibility of mass transfer between their members, which can influence their evolution in many, often unexpected, ways. If the orbit is small enough, one (or both) of the binary members fills its inner common equipotential surface (Roche lobe) and matter flows towards its binary companion. Depending on the responses of the mass-losing star and of the Roche lobe, mass transfer may proceed on a short time scale (thermal or dynamical), decreasing enormously the a priori probability of the system being observed during this phase. On the other hand, mass transfer on longer time scales (e.g., nuclear) can be very long lived, allowing the mass-losing star to remain in hydrostatic and thermal equilibrium. In this case the system has an increased probability of being observed, and indeed several different kinds of binary systems are observed during a mass transfer phase. It is often the case that the presence of a group of binaries is revealed just because of the interaction between its members, as the process of mass transfer can lead to strong emission at wavelengths where single stars are inefficient emitters.

Low-mass X-ray binaries (LMXBs) belong to one of the distinct groups of interacting binaries that would not have been detected as X-ray sources in the absence of mass transfer. A LMXB is believed to consist of a low-mass main-sequence or giant branch star and a compact object, a neutron star or black hole (e.g., van Paradijs 1991). The low-mass star overflows its Roche lobe and matter is accreted by its compact companion. The observational properties of these systems indicates that mass transfer is able to proceed for an appreciably long time and their typical X-ray luminosities correspond to sub-Eddington mass accretion rates ($\sim 0.01\ \dot{M}_{Edd}$ up to $\dot{M}_{Edd}$). The absence of observed companions more massive than $\sim 2\,M_\odot$ has been attributed (van den Heuvel 1975) to their inability to transfer mass in such a mode that the system becomes a persistent long-lived X-ray source. In this paper we will examine the constraints that a system consisting of a neutron star and a normal star must satisfy in order to sustain mass transfer at a rate appropriate to observed LMXBs for an appreciable interval in their evolution. The range of allowed donor masses and orbital separations will be identified.

The motivation for the present study arises from our interest in population synthesis calculations (Romani 1992; Webbink & Kalogera 1994; Iben, Tutukov, & Yungel'son 1995) for the origin of LMXBs. In the course of our studies, we have come to realize that a way of identifying systems observable as LMXBs constitutes a major element of synthesis calculations. The criteria we develop below specifying the parameter space (donor mass and orbital separation) occupied by newly formed LMXBs apply to any evolutionary channel leading to LMXB formation. These criteria also provide initial conditions for studies of their secular evolution.

We shall see below that the parameter space available to nascent LMXBs is circumscribed and divided by a variety of criteria, relating to (i) the thermal and dynamical stability of the donor, (ii) the nature of mass transfer (and especially of systemic mass and angular momentum losses which may accompany it), and (iii) the age of the parent population. The first two types of constraints are addressed in the following section (§ 2). In the course of that discussion we will see that the range in donor mass and radius (and hence, implicitly, orbital period and separation) available to potential LMXBs is extremely limited, if the mass transfer process conserves total mass and orbital angular momentum (§ 2.1). Most normal star-neutron star configurations are unstable to either thermal or dynamical time scale mass transfer, producing mass transfer rates greatly exceeding the Eddington limit. Although these binaries may not appear as X-ray sources, the possibility of extensive mass and angular momentum loss under these circumstances forces us to re-examine the stability issue allowing for super-Eddington losses of mass and angular momentum (§ 2.2). Exploring further the fate of donor stars in systems undergoing thermal time scale mass transfer, we show in the following section (§ 3) that, under suitable circumstances, they may, later in the course of that interaction, (i) recover thermal stability and enter a long phase of slow mass transfer, in which case they may become X-ray luminous; or (ii) develop full-blown dynamical instability. The limitations on possible LMXB donor stars imposed by the age of the parent population are described in § 4. Finally all of these constraints are collected in § 5, where we address some of their implications for the range of structural properties expected among LMXBs, and inferences drawn from the observed populations of LMXBs and binary millisecond pulsars for the evolutionary processes which may create and link them. An Appendix provides details of the analytic expressions used to quantify the criteria and constraints explored in this paper.

## 2. STABILITY OF THE MASS TRANSFER PHASE

The lifetime of any mass transfer phase depends strongly on the behavior of the mass-losing star. When a star loses mass it first responds adiabatically on a dynamical (very short) time scale in an effort to restore its hydrostatic equilibrium, and then on a thermal (longer) time scale in order to regain its thermal equilibrium. In the case that a star cannot remain in hydrostatic (thermal) equilibrium, it becomes unstable against dynamical (thermal) time scale mass transfer. For stars that are unstable to dynamical mass transfer, thermal stability considerations are irrelevant, since the dynamical time scale is orders of magnitude shorter than the thermal time scale. As we shall see below, for a system to qualify as an observable LMXB, it must experience a long-lived X-ray phase, during which mass transfer is slow and the donor is able to remain in hydrostatic and thermal equilibrium.

The ability of the mass-losing star to remain within its Roche lobe depends on its internal structure. In stars with radiative envelopes, the entropy per unit mass (specific entropy) increases rapidly towards the stellar surface. It is exactly this steep entropy profile that suppresses convection, since a fluid element displaced upwards at pressure equilibrium with its surroundings has lower entropy, and hence higher density than the ambient medium. Mass loss on a rapid time scale, such that the star cannot remain in thermal equilibrium, brings to the surface gas with much lower entropy than that of a star of the same instantaneous mass in thermal equilibrium. The envelope of the mass-losing star is therefore denser, and less extended, and its radius smaller. Thus the existence of a radiative envelope and thermal disequilibrium enables the star to track its Roche lobe, and mass transfer proceeds on the time scale characterizing internal energy redistribution in the donor star, namely its thermal time scale (e.g., Webbink 1985). In the case that the star has a deep convective envelope the specific entropy profile is nearly flat (in fact it is slightly negative so that the convection criterion is satisfied) and mass loss results in an overall expansion, because of the reduction in self-gravity holding the star together. Such a star is unable to remain in hydrostatic equilibrium within its Roche lobe and mass transfer then proceeds on the dynamical time scale of the star.

As mass is lost from the donor both the stellar and the Roche lobe radius change, and the characteristics of mass transfer are determined by the interplay between the stellar and Roche-lobe responses. In discussing stability criteria it is convenient to use radius-mass exponents (see, e.g., Webbink 1985), defined as the logarithmic derivative of radius with respect to mass: $\zeta \equiv d\ln R/d\ln M$. The exponents describing the adiabatic and thermal-equilibrium response may be denoted by $\zeta_{ad}$ and $\zeta_{th}$, respectively, and the one describing the response of the Roche lobe by $\zeta_L$. If $\zeta_L < \zeta_*$, where $\zeta_*$ can be either $\zeta_{ad}$ or $\zeta_{th}$, then the star remains in equilibrium because, on losing mass, it contracts more rapidly than its Roche lobe; otherwise mass transfer becomes unstable.

When the mass transfer rate remains below the Eddington critical value, we assume that total mass and orbital angular momentum are conserved. For a neutron star of gravitational mass equal to $1.4\,\mathrm{M}_\odot$ the Eddington mass accretion rate, $\dot{M}_{Edd}$, is of the order of $10^{-8}\,\mathrm{M}_\odot\,\mathrm{yr}^{-1}$. If mass transfer proceeds on a time scale $\tau$, then the mass transfer rate can be approximated by $\sim M_d/\tau$, where $M_d$ is the mass of the donor. A critical donor mass can be estimated for which the characteristic mass transfer rate becomes equal to the Eddington limit, for each of the three time scales that describe the responses of single stars to perturbations of their equilibrium states. These stellar time scales are:

- *dynamical time scale*
  characterizing the rate at which a star recovers its hydrostatic equilibrium after it has been perturbed. It can be estimated by the free fall time scale:
  $$\tau_{dyn} \approx \frac{1}{(G < \rho_d >)^{1/2}} \simeq 5 \times 10^{-5} \left(\frac{M_d}{M_\odot}\right)^{-1/2} \left(\frac{R_d}{R_\odot}\right)^{3/2} \text{ yr} \qquad (1)$$

- *thermal time scale,*
  characterizing the rate at which a star responds when the balance between production and release of energy is perturbed. In other words, it is the time at which the star regains its thermal equilibrium. It can be estimated by the ratio of the thermal energy content (in hydrostatic equilibrium) to the stellar luminosity:
  $$\tau_{th} \approx \frac{G\ M_d^2/R_d}{L} \simeq 3 \cdot 10^7 \left(\frac{M_d}{M_\odot}\right)^2 \left(\frac{R_d}{R_\odot}\right)^{-1} \left(\frac{L_d}{L_\odot}\right)^{-1} \text{ yr} \qquad (2)$$

- *nuclear time scale,*
  characterizing the rate at which principal nuclear fuels are exhausted. Central hydrogen burning is by far the longest nuclear burning phase and the corresponding time scale is:
  $$\tau_{nuc} \approx \frac{M_c \epsilon_N}{L} \simeq 10^{10} \left(\frac{M_d}{M_\odot}\right) \left(\frac{L_d}{L_\odot}\right)^{-1} \text{ yr} \qquad (3)$$
  where $M_c$ is the mass of the stellar core and $\epsilon_N$ is the energy released per gram due to hydrogen burning.

If the donor star has a convective envelope, mass transfer may be driven by angular momentum losses due to a magnetic stellar wind. We may estimate the characteristic time scale as (cf. Rappaport, Verbunt, and Joss 1983):
$$\tau_{MSW} \approx 5 \times 10^6 \left(\frac{M_d + M_{NS}}{M_\odot}\right)^{-2} \left(\frac{M_{NS}}{M_\odot}\right) \left(\frac{R_d}{R_\odot}\right)^{-\gamma} \left(\frac{A}{R_\odot}\right)^5 \text{ yr}, \qquad (4)$$
where $M_{NS}$ is the mass of the neutron star. We have adopted a value of $\gamma = 2$, which Rappaport et al. (1983) find best reproduces the gap in the orbital period distribution of cataclysmic variables.

Magnetic stellar winds alone do not appear capable of driving mass transfer at rates exceeding the Eddington critical rates. Mass transfer on a nuclear time scale becomes super-Eddington for $M_d \gtrsim 4\,M_\odot$. If mass transfer occurs on a thermal time scale it will be super-Eddington if $M_d \gtrsim 0.8\ M_\odot$. In the case that mass transfer proceeds on dynamical time scale, the rate exceeds $\dot{M}_{Edd}$ by many orders of magnitude for all possible donor star configurations.

### 2.1. Limits for Conservative Mass Transfer

Clearly, the character of mass transfer, and thus the viability of candidate LMXB progenitors, depends crucially on the thermal and dynamical stability of the donor star, and secondarily on the time scale for its evolution or for the evolution of the binary orbit. Let us turn, therefore, to quantifying these stability limits and their dependence on the binary system.

We can calculate the limits for critical stability of the mass transfer phase by equating the Roche-lobe radius-mass exponent, $\zeta_L$, to each of the stellar exponents, $\zeta_{th}$ and $\zeta_{ad}$. The radius of the Roche lobe, $R_L$, around a star in a binary system can be expressed as a fraction of the orbital separation:

$$R_L = r_L \cdot A \ , \tag{5}$$

where (Eggleton 1983)

$$r_L = \frac{0.49q^{2/3}}{0.6q^{2/3} \ + \ \ln(1+q^{1/3})} \tag{6}$$

and $q$ is the ratio of the mass of the star to its companion mass. In the case of conservative mass transfer, both total mass and orbital angular momentum are constant. Conservation of total mass gives:

$$\frac{\dot{M}_d}{M_d} = \frac{\dot{q}}{q}\,\frac{1}{1+q} \ , \tag{7}$$

which, combined with conservation of orbital angular momentum, yields:

$$\frac{\dot{A}}{A} = \frac{\dot{q}}{q}\,\frac{2(q-1)}{1+q} \ , \tag{8}$$

where $q \equiv M_d/M_g$, $M_d$ being the mass of the donor and $M_g$ the mass of the gainer. The radius-mass exponent of the Roche lobe then depends only on $q$:

$$\zeta_L = \frac{\mathrm{d}\ln R_L}{\mathrm{d}\ln M_d} \ = \ \frac{4}{3}(2q-1) - (1+q)\ F \ \ , \tag{9}$$

where

$$F \equiv \frac{0.4q \ + \ (1/3)q^{2/3}/(1+q^{1/3})}{0.6q \ + \ q^{1/3}\ \ln(1+q^{1/3})} \tag{10}$$

Hjellming (1989) studied the response of a mass-losing star for a wide range of initial stellar masses ($0.25 - 20.0\,\mathrm{M}_{\odot}$) at various evolutionary stages. Since the stellar response depends on the structure of the donor, which varies with its initial mass evolutionary stage, the adiabatic and thermal equilibrium mass-radius exponents are also functions of the donor mass and radius.

Let us briefly summarize Hjellming's results:

Among stars which have not appreciably evolved away from the ZAMS a reversal in stability hierarchy occurs at $\sim 0.75\,\mathrm{M}_\odot$, where $\zeta_{ad} = \zeta_{th}$. More massive donors are dynamically stable ($\zeta_{ad} \gg 0$), because their envelopes are radiative with a steep entropy profile (e.g., Webbink 1985). The criterion for thermal stability thus imposes the more stringent constraint on the binary mass ratio. In less massive donors the development of convection causes the adiabatic exponent to decrease significantly and the dynamical stability criterion becomes the more stringent one.

The thermal and adiabatic radius-mass exponents of a star are also strong functions of the star's evolutionary state. An example of these variations, for a fairly typical star of mass $2\,\mathrm{M}_\odot$, may be found in Fig. 1, where they are plotted not as a function of time, but as a function of stellar radius, this being the parameter of interest for mass-transfer studies. The monotonic decrease in $\zeta_{th}$ seen there during main-sequence evolution is typical of more massive stars, but is moderated for lower masses. Below $\sim 1.3\,\mathrm{M}_\odot$, $\zeta_{th}$ remains essentially constant through main-sequence evolution. The dramatic decrease in $\zeta_{th}$ of an intermediate-mass star during its subsequent expansion through the Hertzsprung gap is also apparent in Fig. 1, and reflects the tendency of single stars in this mass range to expand toward the giant branch at rates approaching a thermal time scale for larger masses. The range in $\log R$ of this expansion increases dramatically with increasing stellar mass, and so then does the extent to which $\zeta_{th}$ becomes more negative, a strongly destabilizing effect. Conversely, this behavior is strongly moderated at lower mass. As a star finally approaches the base of its giant branch evolution, it develops a surface convection zone which deepens very rapidly, and single stars more massive than $\sim 1.3\,\mathrm{M}_\odot$ undergo momentary contraction. This behavior is inherently stabilizing, with respect to mass transfer, and produces the abrupt increase in $\zeta_{th}$ seen at this point (BGB) in Fig. 1. During the subsequent evolution of a star up the giant branch, its response to mass loss is dominated by its deep convective envelope, and the adiabatic radius-mass exponent, $\zeta_{ad}$, becomes the threshold of interest. In short, the evolutionary expansion of a star is accompanied by diminished inherent stability against thermal time scale mass transfer, until, at the base of the giant branch, the stability criterion abruptly changes to one of dynamical stability.

Given a specific mass for the accreting star, critical donor masses for transition to thermal or dynamical time scale mass transfer can now be found by coupling their stellar responses to those of their Roche lobes. The critical donor masses for a specific gainer mass can be found by coupling the stellar response to that of the Roche lobe around the donor.

For stars on the main sequence the thermal equilibrium radius, $R_{th}$, can be expressed as a function of the initial mass of the donor $M_d$, the central hydrogen fraction at the onset of mass transfer $X_c$, and the decreasing mass of the donor $M_d'$, as mass transfer continues (Eq. A7, A12). The thermal exponent can then be calculated by differentiation of this mass-radius relation. By specifying the gravitational mass of the neutron star $M_{NS}$ and equating $\zeta_{th}$ to $\zeta_L$, we find threshold masses, above which donors become unstable against thermal time scale mass transfer. The corresponding curves of critical donor mass, $M_d$, for $M_{NS}$ equal to 1.2, 1.4, and 1.6 $M_\odot$ are shown in Fig. 2.

For stars crossing the Hertzsprung gap, we use the thermal exponents given by Hjellming (1989) to find the minimum gainer mass needed for stability, $M_{gc}^{HG}$, as a function of the donor mass, $M_d$, and position in the gap (Eq. A13). By equating this minimum gainer mass with $M_{NS}$, we can then find the maximum donor mass consistent with thermal stability (see Fig. 2). The prominent extension of the thermal stability curves to high donor mass just before the base of the giant branch is due to the incipient contraction typical of intermediate-mass stars as they reach the giant branch, a phenomenon which enables more massive donors to track (momentarily) the Roche-lobe radius. We note here that the limiting curves for thermal stability are discontinuous as stars move from the main sequence into the Hertzsprung gap, because between the end of the main sequence and the beginning of the Hertzsprung gap, as we define them, a short-lasting phase of overall contraction intervenes. During this contraction phase, mass transfer cannot be initiated (the donor would already have filled its Roche lobe on the main sequence), and so this phase is of no concern in the present context.

For donors which have evolved beyond the base of the giant branch, thermal stability considerations become irrelevant. The development of deep convective envelopes renders these stars unstable to dynamical time scale mass transfer as soon as a critical mass ratio is exceeded. In fact, the equilibrium radius of such a star can be well approximated as a function of the core mass only (Webbink, Rappaport, & Savonije 1983); since mass loss from the envelope does not influence the growth of the core, the thermal radius-mass exponent $\zeta_{th}$ hovers near zero. We therefore use the adiabatic exponents $\zeta_{ad}$ (Hjellming 1989) to fit the critical gainer mass $M_{gc}^{GB}$ (Eq. A15), and calculate the maximum donor masses consistent with dynamical stability (Fig. 2).

Stars which have evolved beyond the base of the giant branch transfer mass on their nuclear time scale, if they have masses small enough for dynamical stability. However, it is possible that this time scale itself becomes short enough to drive mass transfer at super-Eddington rates. Following the formulation presented by Webbink et al. (1983), we can calculate the limiting stellar radius at which the mass-transfer rate reaches the Eddington limit (lines with heavy dots in Fig. 2). Stars exceeding this maximum radius are then able to maintain hydrostatic equilibrium, but may transfer mass in a non-conservative way. Scrutiny of Fig. 2 shows that giant donors exceed the Eddington limit due to their own nuclear evolution before they become dynamically unstable.

### 2.2. Eddington-Stabilized Mass Transfer

We have assumed until now that super-Eddington mass transfer rates exclude potential candidates for nascent LMXBs. But we should re-evaluate what happens in this case because mass transfer need no longer be conservative, and that fact alone will affect the limits for thermal and dynamical stability.

Although there have been several attempts to model super-Eddington accretion (e.g., Klein, Stockman, & Chevalier 1980; Burger & Katz 1983), a general consensus regarding the physical characteristics and the outcome of this phase does not yet exist. The time-dependent and multi-dimensional character of the process by no means facilitates its modeling. In the absence of a clear physical picture we will assume that the neutron star can accrete matter only up to the Eddington mass-transfer rate ($\dot{M}_{Edd}$) and any excess material lost from the donor is eventually lost from the system carrying away a specific angular momentum equal to the specific orbital angular momentum of the neutron star. Viewed in terms of the energetics of accretion, this idealization appears at least plausible. Formally, the Eddington limit is defined in terms of force balance between radiation and gravity. To eject matter, one must still, at a minimum, provide the local binding energy. However, in the case of a neutron star accreting matter from a non-degenerate donor, the energy released per unit mass accreted is of order $10^5$ times greater than the binding energy at the inner Lagrangian point, so we envision super-Eddington accretion as extremely efficient at expelling excess matter from the accretion flow.

In order to re-evaluate the stability limits in the case of non-conservative mass transfer, we need to evaluate the radius-mass exponent of the Roche lobe. We define the parameter $\beta$ as the fraction of the mass lost from the companion and is accreted onto the neutron star, so:

$$\beta \equiv -\frac{\dot{M}_{Edd}}{\dot{M}_d} , \tag{11}$$

where $\dot{M}_{Edd}$ and $\beta$ are defined to be positive. The remaining fraction, $1-\beta$, of mass lost by the donor is assumed lost from the binary, carrying with it an angular momentum content per unit mass equal to that of the neutron star in its orbit, on the grounds that the matter outflow is driven by accretion energy dissipated mostly within $\sim \beta^{-1}$ radii of the neutron star. The rate of change of $M_d$ and $A$ are then functions of $q$ and $\beta$:

$$\begin{aligned} \frac{\dot{M}_d}{M_d} &= \frac{\dot{q}}{q}\,\frac{1}{1+\beta q} \\ \frac{\dot{A}}{A} &= \frac{\dot{q}}{q}\,\frac{2q^2-(1-\beta)q-2}{(1+\beta q)(1+q)} \end{aligned} \tag{12}$$

We also obtain the Roche-lobe radius-mass exponent for $0 \le \beta \le 1$:

$$\zeta_{L\beta} = \frac{2q^2-(1-\beta)q-2}{1+q} \;+\; (1+\beta q)(\frac{2}{3}-F) \tag{13}$$

where $F$ has been defined in Eq. 10. For $\beta = 1$ we recover the expressions for conservative mass transfer (Eq. 7, 8, 9). However, we have seen above that, where thermal or dynamical limits are violated, the characteristic mass-transfer rates exceed the Eddington limit by orders of magnitude. We are therefore interested in examining the consequences of Eq. (13) in the limit that $\beta \to 0$, and henceforth refer to mass transfer in this mode as "Eddington mass transfer".

Comparison between the Roche lobe exponents for conservative and non-conservative mass transfer shows that as the fraction of mass lost from the system $(1-\beta)$ increases, the Roche-lobe exponent decreases systematically (Fig. 3). This results in a stabilization of mass transfer because more massive stars are able to satisfy the stability criterion ($\zeta_{L\beta} < \zeta_*$). The critical curves corresponding to the extreme case of Eddington mass transfer ($\beta = 0$) are shown in Fig. 4 for different evolutionary stages of the donor and for three different neutron star gravitational masses (1.2, 1.4, and 1.6 $M_\odot$). The limiting curves for other values of $\beta$ (from 0 to 1) lie towards smaller masses as the value of $\beta$ increases and systems move from the case of Eddington mass transfer to the conservative case.

## 3. THE FATE OF THERMALLY UNSTABLE SYSTEMS

### 3.1. The Delayed Dynamical Instability

Beyond the limits just derived, even the Eddington mode cannot stabilize mass transfer against thermal or dynamical time scales.

If a donor has evolved beyond the base of the giant branch, the time scale of mass loss approaches the dynamical time scale, and a large amount of mass is deposited onto the neutron star in an extremely short time, quenching any X-rays from the system and most probably creating a common envelope around it. Even if the spiral-in of the core and the neutron star eventually expels the common envelope, the emerging system will not appear as a LMXB: in its subsequent interaction, gravitational radiation alone will drive super-Eddington mass transfer once again (Pringle & Webbink 1975; Tutukov & Yungel'son 1979). It is conceivable that, in the interim, the system becomes a binary radio pulsar, but the study of this phase is outside the scope of this paper.

On the other hand, donors on the main sequence or in the Hertzsprung gap have radiative envelopes and, although they are able to remain in hydrostatic equilibrium, they drive mass transfer on their thermal time scale. Hjellming (1989) has shown that sustained mass loss on a thermal time scale can eventually lead to dynamical instability. This kind of an instability he labeled as "delayed dynamical instability"; a model calculation of a specific system developing this instability may be found in Webbink (1977). Physically, it is a consequence of the fact that the ambient specific entropy in a star with a radiative envelope rises very rapidly, mass-wise, towards the stellar surface. Most of the volume of the star is filled with this high-entropy, low-density gas of the envelope, but the amount of mass contained in it is very small. As rapid mass transfer continues, the high-entropy gas is lost, and since the star does not have time to relax to thermal equilibrium, the steep entropy profile is stripped away, and low-entropy gas from the interior appears in the surface layer. At this point, the stellar envelope becomes nearly isentropic and tends to expand as more mass is lost. If contraction of the Roche lobe has not abated enough to accommodate this behavior, a dynamical instability develops.

A critical initial mass ratio, $q_{cdd}$, for the delayed dynamical instability can be estimated by finding that $R_L$ curve which is just tangent to the $R_{ad}$ curve (stellar radius in hydrostatic equilibrium but with adiabatic expansion in response to mass loss) for some mass smaller than the initial one (i.e., $\zeta_{L0} = \zeta_{ad}$ at this mass). In making this estimate, we note that in thermal time scale mass loss, thermal relaxation within the donor star occurs primarily in its surface layers, which are immediately lost; the deep interior responds nearly adiabatically. The critical mass ratio, $q_{cdd}$, thus marks the case where the adiabatic expansion of the deep interior can just barely be accommodated within the Roche lobe. The corresponding critical curves in the $\log M_d - \log R_d$ plane are shown in Fig. 5 for three different neutron star gravitational masses. Donors with masses exceeding these limits develop delayed dynamical instability and eventually experience common-envelope evolution.

### 3.2. Survival of Thermally Unstable Systems

The stability criteria that newly interactive normal star-neutron star binaries must satisfy in order ot become LMXBs reveal the existence of a distinct group of systems, which do not immediately become LMXBs upon interaction, but may survive to become LMXBs. This group includes systems with donors first filling their Roche lobes on the main sequence or while crossing the Hertzsprung gap which are not able to maintain thermal equilibrium, but which avoid the growth of a delayed dynamical instability. Since mass transfer rates in these systems are super-Eddington by more than an order of magnitude (and their lifetime is very short), we presume that they do not appear as LMXBs. However, given that Eddington mass transfer may suppress the thermal or dynamical instability, which would otherwise lead to common-envelope formation, we wish to explore the possibility that mass transfer eventually subsides to sub-Eddington rates after a certain amount of mass is lost from the donors, in which case the systems may become LMXBs. For those systems which survive the Eddington mass-transfer phase, we seek as well to identify the area in the parameter space they occupy when they enter a long-lived X-ray phase.

Stabilization of the flow becomes possible mainly because of the behavior of the Roche lobe radius as the mass of the donor decreases. The systems of interest here transfer mass on a thermal time scale, at rates of $\sim 10^{-6}\,\mathrm{M_\odot\,yr^{-1}}$, about two orders of magnitude higher than the Eddington limit ($\sim 10^{-8}\,\mathrm{M_\odot\,yr^{-1}}$ for the assumed neutron star masses). Therefore, to a very good approximation, the total mass lost from the donor is lost from the system as well (Eddington mass-transfer). In this case the Roche-lobe radius at any instant after the onset of mass transfer is given by:

$$\frac{R_L}{R_i} = \frac{r_L(q')}{r_L(q)}\,\frac{1+q}{1+q'}\,\left(\frac{q}{q'}\right)^2\,\exp[2(q'-q)], \tag{14}$$

where $R_i$ is the stellar radius at the onset of mass transfer (equal to the Roche-lobe radius at that time), $q$ is the initial mass ratio equal to $M_d/M_{NS}$ and $q' = q(M_d'/M_d)$. This equation is in agreement with one (Eq. A7) given by Bhattacharya & van den Heuvel (1991) after correcting for two typographical errors in their work: $q \equiv m_1/m_2^0$ instead of $q \equiv m_1/m_2$ and $\exp(2(q_0 - q))$ instead of $\exp(2(q - q_0))$ (Bhattacharya 1995) . The behavior of the Roche lobe radius as the lobe-filling star loses mass is qualitatively the same in Eddington mass transfer case as in conservative mass transfer. As the stellar mass $M_d'$ decreases, $R_L$ decreases as well until it reaches a minimum and starts increasing for even smaller masses. If the Roche-lobe radius of the donor then becomes large enough to exceed the thermal equilibrium radius of the star, the donor may be able to relax to equilibrium, and the system becomes a LMXB.

We can estimate the point at which the donor recovers thermal equilibrium, as follows. For donors that started transferring mass at some time during their main-sequence evolution, the thermal-equilibrium radius $R_{th}$ is given by Eq. (A7, A12). Using this equation and Eq.(14) we can find that donor mass, $M_d'$, at which the Roche lobe radius grows equal to $R_{th}$ and the star is able to restore its thermal equilibrium. Both the parameter space occupied by the initially unstable donors and the area over which these donors appear when they appear as LMXBs are shown in Fig. 6 (regions $\mathrm{MS}_1$ and $\mathrm{MS}_2$, respectively), for a $1.4\,\mathrm{M}_\odot$ neutron star. All systems with donors that start transferring mass on the main sequence eventually survive the rapid mass-transfer phase and become LMXBs. Because shrinkage of the Roche lobe is more severe for more massive donors (Fig. 7), but the equilibrium radii of stripped donors, $R_{th}$, are generally comparable to their initial radii, initially more massive donors lead to the least massive LMXB donors. The region $\mathrm{MS}_2$ is therefore related to region $\mathrm{MS}_1$ by a reflection through the line of marginal stability separating them.

A similar process may occur among donors which start transferring mass while they are crossing the Hertzsprung gap. These stars have already formed dense cores in their interiors, and they are evolving rapidly towards the base of the giant branch. Mass loss even on a thermal time scale has little effect on the rapid evolution of the core, which continues to contract. By the time that the Roche lobe starts expanding, the donor has typically evolved close to the giant branch, and therefore its equilibrium radius is dictated by the core mass at the base of the giant branch (Webbink et al. 1983). A donor will survive this rapid mass-loss phase only if $R_L$ becomes equal to the giant branch radius appropriate to its core mass before the entire envelope is lost. Not all unstable systems satisfy this criterion: Donors more massive than $\sim 2\,\mathrm{M}_\odot$ are stripped of their entire envelopes during thermal time scale mass transfer. The parameter space occupied by systems that do survive eventually and the corresponding area that they populate once they become LMXBs are also shown in Fig. 6 (regions $\mathrm{HG}_1$ and $\mathrm{HG}_2$, respectively). Survivors ($\mathrm{HG}_2$) bear the same relationship to their progenitors ($\mathrm{HG}_1$), in the sense of more massive progenitors yielding less massive survivors, as was the case for the main sequence donors described above. LMXBs formed by thermal time scale mass loss from donors in the Hertzsprung gap have low mass companions ($\lesssim 1\,\mathrm{M}_\odot$) and relatively wide orbits.

## 4. AGE-RELATED LIMITS

Apart from the set of constraints concerning stability of the mass transfer phase, nascent LMXBs must satisfy one additional constraint related to their age. The age of the Galactic disk sets an upper limit on the age of LMXBs in the disk. Therefore, to the extent that LMXB donors survive to the X-ray stage essentially unaltered in mass through any prior phase of binary interaction, their radii must be smaller than the maximum radii that such stars would normally attain in their undisturbed evolution within that time limit, a value which depends on their masses. This constraint is in reality age-dependent only for low-mass donors. More massive stars reach the end of their evolution in a time shorter than the age of the disk and no upper limit on their radii is imposed from age considerations. An absolute lower limit to the radius of the donor star also exists and is dictated by the stellar radius on the Zero-Age Main Sequence (ZAMS) (Eq. A1).

Several estimates of the Galactic disk age exist in the literature, based mainly on studies of old open clusters and they cover a relatively wide range of values, $6-10\times10^9$ yr (e.g., Demarque, Green, & Guenther 1992; Hobbs, Thorburn, & Rodriguez-Bell 1990). we have explored the dependence of the LMXB population on the Galactic disk age, and have calculated the limiting radii for three different values of the Galactic disk age: $T_{10}=0.8, 1.0, 1.2$, where $T_{10}$ is the age in units of $10^{10}$ yr (see Fig. 8). In deriving these limits we have used the evolutionary calculations of Schaller et al. (1992).

## 5. DISCUSSION

The structural and evolutionary constraints which we have identified above define the region in donor star masses and orbital separations (or orbital periods) which neutron star-normal star binary systems may inhabit when they first reach interaction. We expect only a small part of this region to produce LMXBs directly upon the onset of mass transfer. In a much wider range of parameters, mass transfer develops on a dynamical time scale, almost certainly leading to a new episode of common envelope evolution (which the normal star donor can only possibly survive if it has developed a hydrogen-exhausted, and probably degenerate core); or on a thermal time scale, which may develop into dynamical time scale mass transfer, stabilize at a slower (possibly becoming an LMXB) in the course of mass transfer, or continue unabated until the donor is stripped to a degenerate core. Under conditions in which they occur in the present context, both dynamical and thermal time scale mass transfer invariably proceed at strongly super-Eddington rates, but super-Eddington mass transfer may also occur even in the absence of these instabilities if the donor's intrinsic evolutionary time scale is short enough.

These various evolutionary channels are summarized in Fig. 9a, which provides an explicit division of the donor star mass-radius plane into relevant regimes. The Roche lobe formalism (Eq. 5 and 6) permits a straightforward mapping of the donor star mass-radius plane (Fig 9a) into one of donor star mass vs. orbital period (Fig. 9b). For simplicity, we show only the boundaries for a population of age $1.0 \times 10^{10}$ yr, with neutron star gravitational masses of $1.4\,M_{\odot}$; dependencies of the various boundaries on age and neutron star mass are illustrated in the preceding figures (Fig. 2, 4, 5, and 8) in the context of the criteria which define them. Different regions in Fig. 9a are labeled according to the character of their mass transfer, using obvious notation: $D$ – *d*ynamical time scale; $T$ – *t*hermal time scale, $S$ – *s*table dynamically and thermally (hence, slow mass transfer, on a nuclear or angular momentum loss time scale). Hybrid cases are represented by a dual notation: $TS$ – *t*hermal mass transfer evolving into *s*table, slow flow; and $TD$ – *t*hermal mass transfer developing into delayed *d*ynamical instability. In the case of pure slow mass transfer, we are further able to distinguish between systems which initially transfer mass at super-Eddington rates ($S^E$) from those initially sub-Eddington ($S_E$). Detailed evolutionary calculations will be needed to determine whether the stable phase of the hybrid case $TS$ proceeds at sub- or super-Eddington rates. As noted above, thermal and dynamical phases are invariably super-Eddington.

The dual questions of the X-ray observability of sources accreting at super-Eddington rates, and of their evolutionary fate, are clearly central to problems of population syntheses for LMXBs and for binary millisecond pulsars (MPSRs). We are not in a position to provide rigorous theoretical answers to these questions, although our expectation in regard to X-ray observability is that even modestly super-Eddington mass transfer rates may quench X-ray emission by reprocessing and thermalization of radiation in the immediate vicinity of the neutron star, an effect greatly amplified at low X-ray energies by the presence of many high ionization potential species in the relatively metal-rich accretion flow of the Population I binaries modeled here.

Apart from any theoretical considerations, however, evidence that systems can survive a phase of super-Eddington mass transfer comes from observed systems. The existence of MPSRs in circular orbits with periods exceeding the $\sim 100$ d (B0820+02, B1310+18, B1620-26, B1800-27, and B1953+29) indicates that their progenitors (neutron star-giant star systems) must have experienced a mass transfer phase. However, we find that giant donors transferring mass at sub-Eddington rates appear only in systems with $P \lesssim 1.5$ d (see Fig. 9b), not wide enough to leave systems with $P > 100$ d. These longer-period millisecond binary pulsars must have originated in systems in which the donors drove super-Eddington mass transfer ($S^E$ on the giant branch). Time-dependent calculations of the evolution of these systems will prove to be significant in understanding the origin of the long-period MPSRs. However, our stability analysis strongly suggests that these systems arrived at the long-period MPSRs stage without ever having been long-lived detectable X-ray sources. They may justifiably be excluded in partial resolution of a possible LMXB death rate - millisecond pulsar birth rate discrepancy (Kulkarni & Narayan 1988; Coté & Pylyser 1989;

Lorimer 1995).

Further evidence for the survival of systems following a super-Eddington phase comes from the ultra-short-period (685 s) LMXB 4U 1820-30, which is a neutron star probably accreting from a very low-mass ($\sim 0.07\,M_{\odot}$) degenerate companion (Stella, Priedhorsky, & White 1987; Verbunt 1987; Rappaport et al. 1987). All viable formation channels leading to such a binary posit that the donor originated as the core of a giant branch star. Even the least massive star to reach the giant branch within a Hubble time has developed a helium core mass of $\sim 0.15\,M_{\odot}$, which is therefore the minimum initial mass of a degenerate donor. Gravitational radiation alone will drive mass transfer from a degenerate star of this mass at super-Eddington rates (Pringle & Webbink 1975; Tutukov & Yungel'son 1979;), implying that 4U 1820-30 is the survivor of a super-critical mass transfer phase.

There exist a small number of very short-period LMXBs (4U 1626-67 and 4U 1916-05) which for structural reasons (Rappaport & Joss 1984) are believed to contain hydrogen deficient donors. Their composition implies that the donor stars had their nuclear evolution terminated abruptly late in their main sequence evolution by the onset of rapid mass transfer. In all probability, therefore, these systems originated from the region on the main sequence labeled $TS$ (Fig. 9a,b), and are likely to be survivors also of a super-Eddington phase.

Orbital periods have been measured for 20 LMXBs and in 6 of them they are in excess of 20 h (van Paradijs 1995). These long periods imply that mass transfer is driven by nuclear evolution of the donors, which have already reached the giant branch (Fig. 9b). We have shown that, of the region in parameter space occupied by donors transferring mass at sub-Eddington rates, only a small part is occupied by evolved stars. The extent of this region is very sensitive to the age of the parent population, but in any case seems unlikely to account for the observed incidence of systems with giant donors. On the other hand, we consider it likely that the true incidence of evolved donors is lower than the figures above would indicate, since selection effects favor the optical identification of systems with evolved, hence luminous, donors. In any event, survival through a super-Eddington phase opens avenues for the creation of these systems either from stars initially reaching mass transfer on the giant branch, and stabilized by super-Eddington mass transfer ($S^E$), or from those surviving thermal time scale mass transfer initiated while they were crossing the Hertzsprung gap.

In this paper we have restricted our study to systems containing accreting neutron stars only. However, measurements of dynamical masses of compact objects in some LMXBs suggest that they are black holes. The estimated black hole masses span a range of values much wider than that of NS-masses, from $\sim 4$ up to $10\,M_{\odot}$ (Cowley 1992; Wijers 1995).

We expect that an increase of the gainer mass will result in an extension of the parameter space occupied by stable donors towards higher masses, primarily on the main sequence and the giant branch, and to a much smaller extent in the Hertzsprung gap. However, the systems populating this extension transfer mass at super-Eddington rates, even for black hole accretors, since their nuclear time scales are much shorter. Because of this same sensitivity of nuclear time scale on mass, the expansion of *sub*-Eddington parameter space is almost certainly much more modest.

We have outlined here limits for conservative and non-conservative mass transfer under specific idealized conditions which tend to favor stability. In reality, we expect additional mass and angular momentum losses to come into play. In the case of super-Eddington mass transfer, for example, we have ignored any contribution to the angular momentum of the matter outflow arising from its rotation with respect to the accreting neutron star. Such additional angular momentum losses tend to de-stabilize donors. We, therefore, regard the conditions assumed here as the most favorable for stability of the donor stars.

Finally, it is important to point out that all the constraints that nascent LMXBs must satisfy are independent of the specific evolutionary path that their progenitors have followed. The relative populations of the different groups of systems discussed here will depend not only on the relative sizes of the areas in the parameter space that they occupy in Fig. 9b, but also on their relative birth rates and lifetimes. The latter characteristics can be calculated via population synthesis calculations, for which the knowledge of the properties of nascent LMXBs is necessary, and which we will explore in subsequent papers.

It is a pleasure to thank F. K. Lamb and D. Psaltis for stimulating discussions. This work is supported by National Science Foundation under grant AST92-18074.

## A. ANALYTIC EXPRESSIONS FOR STELLAR PARAMETERS

A set of analytic approximations have been employed in the calculation of the limiting curves on the parameter space of nascent-LMXB donors. The stellar models used in all cases assumed solar composition. The units used throughout are: masses in $M_\odot$; radii and orbital separations in $R_\odot$; time in yr. Natural logarithms are written as "ln", decimal logarithms as "log", and the arguments of trigonometric functions are in radians.

We have used stellar models by Schaller et al. (1992), Eggleton (unpublished) and Webbink (unpublished) to approximate the stellar radius of the donor as a function of donor mass $M_d$ on the zero-age main sequence (ZAMS):

$$\begin{aligned} R_{ZAMS} &= \left(10^{-15z_1} + 10^{-15z_2}\right)^{-1/15} && \log M_d \geq -0.0909 \\ &= \left(10^{z_3} + 10^{z_4}\right)^{-0.06} && -0.0909 > \log M_d \geq -0.1974 \\ &= 10^{z_5} && -0.1974 > \log M_d \geq -1.05 \end{aligned} \quad \text{(A1)}$$

where

$$\begin{aligned} z_1 &= 2.359(\log M_d)^2 + 1.195(\log M_d) - 6.864 \times 10^{-2} \\ z_2 &= 0.547(\log M_d) + 4.421 \times 10^{-2} \\ z_3 &= -29.17(\log M_d)^2 - 24.08(\log M_d) + 0.68 \\ z_4 &= -5.83(\log M_d)^2 - 6.58(\log M_d) - 0.33 \\ z_5 &= -0.126 + 0.715(\log M_d) + 0.016 M_d^{1/2} \sin(8(\log M_d) + 2.4) \end{aligned} \quad \text{(A2)}$$

on the terminal main sequence (TMS):

$$R_{TMS} = \left(10^{-15t_1} + 10^{-15t_2}\right)^{-1/15} \qquad 1.4 \geq \log M_d \geq -0.2 \quad \text{(A3)}$$

where

$$\begin{aligned} t_1 &= 2.824(\log M_d)^2 + 1.325(\log M) + 0.115 \\ t_2 &= 0.112(\log M_d)^2 + 0.467(\log M) + 0.357 \end{aligned} \quad \text{(A4)}$$

and at the base of the giant branch (BGB):

$$R_{BGB} = \left(10^{-4b_1} + 10^{-4b_2}\right)^{-1/4} \qquad 20.0 \geq M_d \geq 0.8 \quad \text{(A5)}$$

where

$$\begin{aligned} b_1 &= 1.208(\log M_d)^2 + 1.207(\log M) + 0.242 \\ b_2 &= 1.112(\log M_d)^2 - 0.235(\log M) + 1.343 \end{aligned} \quad \text{(A6)}$$

Stellar models by Horn, Kříž, & Plavec (1970), Hjellming (1989) and Webbink (unpublished) have been used to approximate the thermal equilibrium radius $R_{th}$ of a mass-losing main sequence star as a function of present mass $M_d'$, initial mass $M_d$, and central hydrogen fraction $X_c$. For masses $M_d'$ close to the initial one $M_d$ and for $M_d \lesssim 5\,\mathrm{M}_\odot$ the thermal equilibrium radius can be approximated by:

$$\begin{aligned} R_{th} &= R_o \left(\frac{M_d'}{M_d}\right)^{M_d[X_c(1-X_c)-0.25]} & X_c \leq 0.5 \\ &= R_o & X_c > 0.5, \end{aligned} \tag{A7}$$

where $R_o$ is the radius of a star with mass equal to $M_d'$, which has evolved at constant mass away from the ZAMS and its central hydrogen fraction is $X_c$. This radius is calculated under the assumptions that: (i) the hydrogen fraction $X_c$ varies parabolically with time:

$$X_c = -0.427\left(\frac{t}{t_{MS}}\right)^2 - 0.273\left(\frac{t}{t_{MS}}\right) + 0.7, \tag{A8}$$

where $t_{MS}$ is the main sequence lifetime, and (ii) the stellar radius $R$ at any time is given by:

$$R = R_{ZAMS}\left(1 - \frac{t}{t_{ev}}\right)^{-0.28} \tag{A9}$$

where

$$\log t_{ev} = 0.894(\log M_d)^2 - 3.601 \log M_d + 10.111 \tag{A10}$$

The main sequence lifetime $t_{MS}$ is defined implicitly by equating Eq. A3 and Eq. A9 and solving for $t$. The behavior of $R_{th}$ changes beyond an inflection point which occurs at locations $(\log M_{ip}, \log R_{ip})$ defined by the crossing point between Eq. (A7) and:

$$\log R_{ip} = 0.8 \log M_{ip} - \frac{0.015}{\log M_{ip} - 0.09} + 0.55 \tag{A11}$$

Beyond this inflection point $R_{th}$ becomes:

$$\log R_{th} = \log R_{ip} + 0.95(\log M_d' - \log M_{ip}) \qquad M_d' < M_{ip} \tag{A12}$$

For even smaller masses ($M_d' \lesssim 0.8\,\mathrm{M}_\odot$) the behavior of $R_{th}$ changes again, but this change occurs at masses which are out of the range of interest to us.

Using calculations presented by Hjellming (1989) we have approximated the critical gainer mass $M_{gc}^{HG}$ for donor thermal stability in the Hertzsprung gap as a function of the donor mass $M_d$ ($M_d \lesssim 2\,\mathrm{M}_\odot$) and a normalized variable $\hat{r}_{HG}$, which describes the position of the star in the Hertzsprung gap:

$$\begin{aligned} M_{gc}^{HG} = \Big[&\left((-9.5M_d + 11.875)\hat{r}_{HG} + (9.4626M_d - 10.8)\right)^{-20} + \\ &\left((-3.8M_d + 14)\hat{r}_{HG}^2 + (3.3M_d - 3)\right)^{-20}\Big]^{-1/20}, \end{aligned} \tag{A13}$$

where

$$\hat{r}_{HG} = \frac{\log R - \log R_{TMS}}{\log R_{BGB} - \log R_{TMS}}. \tag{A14}$$

Similarly, the critical gainer mass $M_{gc}^{GB}$ for dynamical stability of the donor on its giant branch for $M_d \lesssim 2\,\mathrm{M}_\odot$ can be approximated by:

$$M_{gc}^{GB} = (1.55 M_d - 0.28)\left[1 - \exp\left(\frac{\hat{r}_{GB}}{a} + b\right)\right], \tag{A15}$$

where

$$\begin{aligned}
a &= -0.245 M_d + 0.46 && M_d \leq 1.55 \\
&= 0.08025 && M_d > 1.55 \\
b &= -2.84 M_d + 1.86 && M_d \leq 1.25 \\
&= -2.52 M_d^2 + 9.26 M_d - 9.325 && M_d > 1.25 \\
\hat{r}_{GB} &= \frac{\log R - \log R_{BGB}}{\log R_{BGB}}. && \quad\quad (A16)
\end{aligned}$$

## Figure Captions

Fig. 1— Radius-mass exponents of a $2\,M_\odot$ star throughout its evolution (data taken from Hjellming 1989). Adiabatic and thermal exponents are indicated by open and filled circles, respectively.

Fig. 2— Radius-mass diagram for donor stars. Heavy solid lines mark the loci of zero-age main sequence stars (ZAMS), terminal main sequence stars (TMS), and stars at the base of the giant branch (BGB). Thin lines mark critical (maximum) donor star masses for three choices of neutron star gravitational masses. Critical masses on the main sequence (between ZAMS and TMS) and in the Hertzsprung gap (between TMS and BGB) are limits for thermal stability, those on the giant branch (above BGB) for dynamical stability. Donors to the right of these lines are unstable, while those to the left are stable. Also shown (lines with heavy dots) are critical (maximum) radii beyond which nuclear evolution drives super-Eddington mass transfer. Critical masses and radii all assume conservative mass transfer.

Fig. 3— Roche lobe radius-mass exponent as a function of mass ratio $q$ for different fractions of mass lost from the system $\beta$. The value of $\beta = 1$ corresponds to conservative and $\beta = 0$ to "Eddington" mass transfer.

Fig. 4— Radius-mass diagram as in Fig. 2, but in the limit of "Eddington" mass transfer.

Fig. 5— Critical (minimum) donor masses for the development of a delayed dynamical instability for three different neutron star masses.

Fig. 6— Areas in the parameter space occupied by donors initially unstable to thermal time scale mass transfer ($MS_1$ and $HG_1$) and those they occupy when they emerge from rapid mass loss ($MS_2$ and $HG_2$, respectively). A neutron star gravitational mass $M_{NS} = 1.4\,M_\odot$ is assumed.

Fig. 7— Roche lobe radius as a function of decreasing mass ratio $q$ for two values of initial mass ratios $q_i$ in the limit of "Eddington" mass transfer.

Fig. 8— Maximum donor radii for three different ages, $T_{10}$ (in units of $10^{10}$ yr), of the Galactic disk.

Fig. 9— The complete set of limits on the properties of (a) donors and (b) binary systems for the specific case of $M_{NS} = 1.4\,M_\odot$ and $T_{10} = 1.0$.